\documentclass[aps,prb,twocolumn,showpacs,superscriptaddress,groupedaddress]{revtex4}
\usepackage[compat=1.0.0]{tikz-feynman}
\usepackage{amssymb}
\usepackage{bm}
\usepackage{amsmath}
\usepackage{graphicx}
\usepackage{epstopdf}
\usepackage{subfigure}
\usepackage{natbib}
\usepackage{epsfig}
\usepackage{amsfonts}
\usepackage{mathrsfs}
\usepackage[toc,page,title,titletoc,header]{appendix}
\usepackage[colorlinks,linkcolor=blue,citecolor=blue,anchorcolor=blue]{hyperref}
\usepackage{dsfont,amsthm,amsbsy}

\usepackage{fancyhdr}

\begin{document}

\newcommand{\com}[1]{[{\bf  \color{blue}{Comments: #1}}]}

\title{Charge Density Waves in a Quantum Plasma}
\author{Zhaoyu Han}
 \affiliation{Department of Physics, Stanford University, Stanford, California 94305, USA}

\author{Shiwei Zhang}
\email{szhang@flatironinstitute.org}
 \affiliation{Center for Computational Quantum Physics, Flatiron Institute, New York, New York 10010, USA}
 \affiliation{Department of Physics, The College of William and Mary, Williamsburg, Virginia 23187, USA}

\author{Xi Dai}
\email{daix@ust.hk}
 \affiliation{Physics Department, Hong Kong University of Science and Technology, Hong Kong, China}

\date{\today}

\newcommand{\co}[1]{[{\bf  \color{red}{Comment: #1}}]}

\begin{abstract}
We analyze the instability of an unpolarized uniform quantum plasma consisting of two oppositely charged fermionic components with varying mass ratios, against charge and spin density waves (CDW's and SDW's).
Using density functional theory, we treat 
each component 
with the local spin density approximation and a rescaled exchange-correlation functional. 
Interactions between different components are treated with a mean-field approximation.  
In both two- and three-dimensions, 
we find leading unstable CDW modes in the second-order expansion of the energy functional, which would induce the transition to quantum liquid crystals.
The transition point and the length of the wave-vector are computed numerically. Discontinuous ranges of the wave-vector are found for different mass ratios between the two components, indicating exotic quantum phase transitions. 
Phase diagrams are obtained and a scaling relation is proposed to generalize the results to two-component fermionic plasmas with any mass scale. 
We discuss the implications of our results and directions for further improvement in treating quantum plasmas.
\end{abstract}

\maketitle

\section{Introduction}
Plasma, as one of the four fundamental states of matter, can be  generally understood as a mixture of roaming ions, whose behavior is usually dominated by collective effects mediated by the electromagnetic force. Past studies have primarily focused on the classical or semi-quantum region, where at least one of the components is not fully treated with quantum mechanics. For example, calculations with the coupled electron-ion Monte Carlo (MC) method typically neglect  ionic exchange interaction~\cite{BookCEIMC, CEIMC2010}, and path integral MC or molecular dynamics methods often do not include a full description of quantum statistics~\cite{MD1998, PIMC2001, DPIMC2003A, DPIMC2003B, PIMC2005, BookPIMC}. Those simplified calculations considered not only the computational challenges and expenses of a complete treatment of full quantum effects, but also the relative rarity of situations where the plasma is dense and cold enough so that quantum effects dominate the behaviors of all component. Such systems, however, can be found in the interior of giant planets or white dwarf stars, and in the world of condensed matter physics. For example, in semiconductors, the effective particles and holes introduced by the electronic band structure could play the roles of the two different types of ions~\cite{ehliq,ex1,ex3}, for which the behavior must be understood with quantum theories. (See, e.g. Ref.~\cite{QMCeh} for case studies of equal masses.) Recently, increasing attention has focused on nuclear quantum effects~\cite{nuclear-quantum-JCP-special-2018}.

A particular kind of plasma that can be viewed as of extremely large mass ratio, the electron gas in the background of positively charged jellium, is one of the most fundamental models in many-body physics and has been extensively investigated~\cite{uni,liq}. Density Functional Theory (DFT) calculations~\cite{KS19651,KS19652,martin_2004} rely on the correlation energies of the electron gas as a foundation. Because of both analytical and numerical challenges, the phase diagram of this model remains incomplete. The intermediate phases between the high-density limit and the opposite limit, which are the uniform liquid phase and the Wigner crystal phase respectively, are uncertain~\cite{Wigner,2DWC,DFTWig3D,DFTWig2D}. Quantum MC (QMC)~\cite{cep80,GS2D,2DXC,att02,att03,cep79,cep93,ort94,ort97,ort99,Zong,drum04,X05,X06,cla09,drum13} calculations, which have provided the parametrization for the correlation energies to serve as the basis for most modern DFT calculations, are the most sophisticated numerical treatment. However, one can still be limited by the candidate structure or accuracy (e.g. from the fixed-node approximation~\cite{cep80} with the trial wave function), finite-size effects, and incommensurability with the true ground state structure.  Many Hartree-Fock (HF)~\cite{2DHF,need03,zh08,hf08,3DHF} calculations indicate possible additional phases of magnetic and charge order, but the relevance of these predictions  to the actual many-body ground state is difficult to establish because of the crude nature of the approximation.

In this paper, the quantum limit (i.e. at high density and zero temperature) of a two-component plasma is investigated at all mass ratios, by means of DFT within the framework of local spin density approximation (LSDA). Neglecting the correlation effects between the two components beyond electrostatics (Hartree), we calculate the ground-state energy as functional of their density distributions. An analysis of second-order expansion of the energy functional shows that the unpolarized uniform liquid state is unstable against CDWs with infinitesimal amplitudes at certain densities, which could eventually lead to the formation of (smectic) quantum liquid crystals~\cite{Cry,ITO2004148,ITO2007107,Frad12}. We further find discontinuities in the relation between the mass ratio and the magnitude of the leading unstable wave-vector for  both two- (2D) and three-dimensional (3D) cases. These discontinuities may indicate exotic quantum phase transitions between crystalline phases with different structures. The ground-state phase diagrams are concluded and partially conjectured for both 2D and 3D. A simple scaling relation generalizes these results to all mass scales.

\if
{More careful calculations was carried out at small ion mass limit, where the system is equivalent to the jellium model. Intermediate states, as conjectured in ref.[jp], seem to be one-, two- and three-dimensional Wigner crystal as $r_s$ increases. Different from previous LSDA works, our calculation adopts the latest exchange-correlation energy fitting, which significantly modified the magnetic dependence in the intermediate density region.}

\begin{align}
r_s=\left\{\begin{aligned}
(\frac{3}{4\pi\rho_0})^{1/3} \ \ \ D=3 \\
(\frac{1}{\pi\rho_0})^{1/2} \ \ \ D=2\end{aligned}\right.
\end{align}
\fi

\section{Methods}
The system we consider here consists of positive and negative fermionic ions. They are of the same number $N$, equally charged with unit electron charge, and confined in a $D$-dimensional volume $V$. For convenience, we use Wigner-Seitz radius $r_s$, which is the radius of a sphere containing one electron, to parametrize the number density $\rho_0=N/V$. All the quantities, operators and equations are in atomic units and subscripted by $p,n$ for the two positively and negatively charged components respectively. The total Hamiltonian for the many-body system reads:
\begin{align}
\mathcal{H}_{\text{total}} = & \sum_{a=p,n}\left(- \sum^N_{i}\frac{\bm{\nabla}_{a,i}^2}{2 m_a} + \sum^N_{i<j} \frac{1}{|\bm{r}_i^a-\bm{r}_j^a|}\right) \nonumber \\&- \sum^N_{i,j} \frac{1}{|\bm{r}_i^p-\bm{r}_j^n|}
\end{align}
Further we will simply use $\gamma\ge1$ to denote the mass ratio between the heavier component and the lighter one, and $m^*$ to represent the larger mass and thus the mass scale, since reversing the signs of the charges does not affect the physics here. 

The main assumption of our treatment is the neglect the quantum correlation between two components. It is equivalent to separating the wave-functions of different components, which is known as Born-Oppenheimer approximation and has been widely adopted in molecular physics studies. The approximation can at least be partially justified at large mass ratio $\gamma$, by recognizing the difference between the time scales of the two components’ motions. We will further discuss the effect of recovering such correlation in Sec.~\ref{dis}.

Then the two subsystems can be viewed independent, except for the local external potentials provided by the other, which arise from the Hartree part of the interaction between them.
The two systems can thus be treated separately with DFT. 
In the high density (low $r_s$) region near the quantum limit, the plasma favors a near-uniform density distribution due to the dominance of the kinetic energy, whose strength is $\propto r_s^{-2}$ overwhelming the $\sim r_s^{-1}$ interaction. Under such circumstance, LSDA can be feasibly applied. 
As we will further discuss below, the reliability of LSDA, both in the sense of 
the accuracy of the functional as fitted from QMC results and, more importantly, as an approximation applied to our many-body Hamiltonian,
is uncertain and will require further validation. Especially in the more
strongly correlated regime, with larger $r_s$ for instance, there can 
be a breakdown.


Assuming $\bm{\rho}^a=(\rho^a_{\uparrow},\rho^a_{\downarrow})$ and $\rho^a = \rho^a_{\uparrow}+\rho^a_{\downarrow}$ represent the (up-, down-) spin and the total density of component $a=p,n$, and defining $\bm{\rho}=(\bm{\rho}^p(\bm{r}),\bm{\rho}^n(\bm{r})) = (\rho^p_{\uparrow},\rho^p_{\downarrow}, \rho^n_{\uparrow},\rho^n_{\downarrow})$, the total ground-state energy as a functional of these density distributions can be written as the sum of the two components' kinetic, exchange-correlation energies and the Hartree energy of the whole system~\cite{KS19651,KS19652}:
\begin{align}\label{functional}
E[\bm{\rho}] = & \sum_{a=p,n}\left( T^a[\bm{\rho}^a] + E^a_{\text{xc}}[\bm{\rho}^a]\right) + E_{\text{Hartree}}[\rho^p,\rho^n],
\end{align}
where $T^a[\bm{\rho}^a]$ is the ground-state energy of an auxiliary non-interacting system with the same density distribution $\bm{\rho}^a$, $E^a_{\text{xc}}[\bm{\rho}^a]=\int \mathrm{d}\bm{r} \rho^a\epsilon^a_{\text{xc}}(\bm{\rho}^a)$ within LSDA, and the Hartree term reads:
\begin{align}
& E_{\text{Hartree}}[\rho^p,\rho^n]\nonumber\\=&\frac{1}{2}\int \frac{\left(\rho^p(\bm{r})-\rho^{n}(\bm{r})\right)\left(\rho^p(\bm{r}')-\rho^{n}(\bm{r}')\right) }{|\bm{r}-\bm{r}'|} \mathrm{d}\bm{r}\mathrm{d}\bm{r}',
\end{align}
which couples the two components of the plasma. We acquire each component's exchange-correlation energy by scaling electron's using the relation in appendix.

It is obvious that the unpolarized uniform solution $\bm{\rho}(\bm{r})=\frac{1}{2}(\rho_0,\rho_0,\rho_0,\rho_0)$ is always a stationary point of this functional, since it satisfies the Kohn-Sham equations derived from the stationary condition $\delta E_0[\bm{\rho}] = 0$. At the high density limit ($r_s\rightarrow0$), this solution is indeed stable. While it is well-known that, at low densities (large $r_s$), 
the ground state of 
the jellium model of electron gas
($\gamma\rightarrow \infty$)
is a Wigner crystal in both two- and three-dimensional cases, explicitly breaking translation symmetry. To investigate the possible symmetry breaking point as a function of $r_s$, we expand the energy functional to the second-order of an arbitrary density fluctuation $\delta \bm{\rho}(\bm{r}) = \rho_0 \sum_{\bm{q}} \delta \bm{\rho}_{\bm{q}} e^{-\mathrm{i}2k_F\bm{q}\cdot\bm{r}}$ around the uniform solution, where $k_F$ denotes the magnitude of the Fermi wave-vector of a non-interacting fermionic system with the same density $\rho_0$,
so that the wave vector $\bm{q}$ is defined in units of $2k_F$. 

We can then directly expand the functional to the second order and transform to momentum space. This expansion can be written into the form of a sum of $4\times 4$ matrices $\mathcal{H}_q$ contracting with density fluctuations $\delta \bm{\rho}_{\bm{q}}$ over all wave vector $\bm{q}$'s:
\begin{align}
\delta^2 E &= \sum_{\bm{q}} \delta \bm{\rho}_{\bm{q}}^{\dagger} \mathcal{H}_q \delta \bm{\rho}_{\bm{q}} \nonumber  
\end{align}
where
\begin{equation} \label{eq3}
\mathcal{H}_q=\frac{N\rho_0}{2}\left(     
  \begin{array}{cc}   
    A+B^p+1/\chi^p & -A\\
    -A & A+B^n+1/\chi^n
  \end{array}
\right)             
\end{equation}

The spin blocks read ($\alpha,\beta=\updownarrow$):
\begin{align}
A_{\alpha\beta} & \equiv h =\left\{\begin{aligned}
\frac{\pi}{k_F^2 q^2} \ \ \ D=3 \\
\frac{\pi}{k_F q} \ \ \ D=2\end{aligned}\right. , \\
B^a_{\alpha\beta} &= \frac{\partial^2 \left(\rho \epsilon^a_{\text{xc}}(\rho_{\uparrow}, \rho_{\downarrow})\right)}{\partial \rho_{\alpha}\partial \rho_{\beta} } {\big |}_{\rho = \rho^a_0},
\end{align} 
which represent the Hartree and exchange-correlation energy variations. We mention 
that the ions' exchange-correlation energy  per-particle, $\epsilon^a_{\text{xc}}$, can be acquired by applying the scaling relation on 
the QMC result for the electron gas. $\chi^a_{\alpha\beta} = \delta_{\alpha\beta} \chi^a_0$ where the static Lindhard function
\begin{align}
\chi_0^a =\left\{\begin{aligned}
\frac{m_a k_F}{4\pi^2} \left(1+\frac{1-q^2}{2q}\ln{\bigg |}\frac{1+q}{1-q}{\bigg |}\right) \ \ \ D=3 \\
\frac{m_a}{2\pi} \left(1-\Theta(q-1)\sqrt{1-1/q^2}\right) \ \ \ D=2\end{aligned}\right.
\end{align}
is the linear response function of the non-interacting fermionic gas~\cite{lin}. 
Its reciprocal evaluates the second variation of the kinetic energy functionals~\cite{HK1964,Perdew}. 

The diagonalizition of $\mathcal{H}_q$ gives two CDW modes' eigen-energy $\lambda_{\text{CDW}}^{\pm}$ and two 
SDW modes' eigen-energy $\lambda_{\text{SDW}}^{a= p, n}$ on each wave vector $\bm{q}$. They are:
\begin{align}
\lambda_{\text{SDW}}^a&=\frac{N\rho_0}{2}\left(2/\chi_0^a+B_{\uparrow\uparrow}^a-B_{\uparrow\downarrow}^a\right), \\
\lambda_{\text{CDW}}^{\pm}&=\frac{N\rho_0}{4}\left(u^p+u^n+4h\pm\sqrt{16h^2+(u^p-u^n)^2}\right),
\end{align}
where we define ($a=p,n$):
\begin{align}
u^a\equiv2/\chi_0^a+B_{\uparrow\uparrow}^a+B_{\uparrow\downarrow}^a.
\end{align}
The corresponding eigenvectors are:
\begin{align}
\bm{\rho}^{p,n}_{\text{SDW}} &= \left(1,-1,0,0\right)/\sqrt{2},\  \left(0,0,1,-1\right)/\sqrt{2}, \\
\bm{\rho}^{\pm}_{\text{CDW}} &= \left(v_{\pm},v_{\pm},1,1\right)/\sqrt{2(1+v_{\pm}^2)},
\end{align}
where
\begin{align}
v_{\pm}\equiv-\frac{(u^p-u^n)\pm\sqrt{16h^2+(u^p-u^n)^2}}{4h}.
\end{align}

With the decrement of density, the first eigenenergy approaching zero gives a leading unstable mode towards the deformation of the unpolarized uniform state.

\section{Results}

\begin{figure*}[t!] 
\subfigure[]{\label{m24}\includegraphics[width=0.3\linewidth]{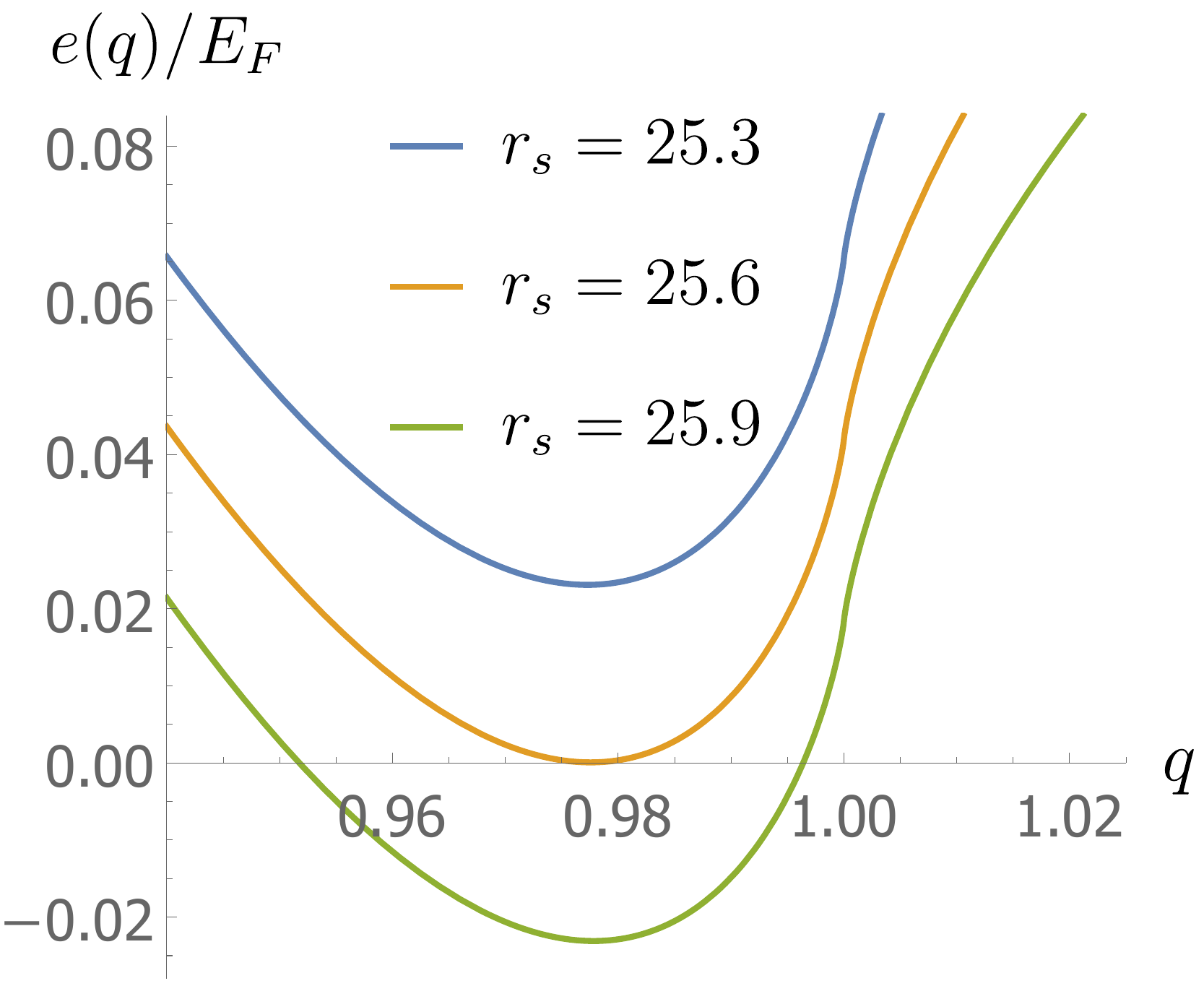}}
\subfigure[]{\label{m240}\includegraphics[width=0.3\linewidth]{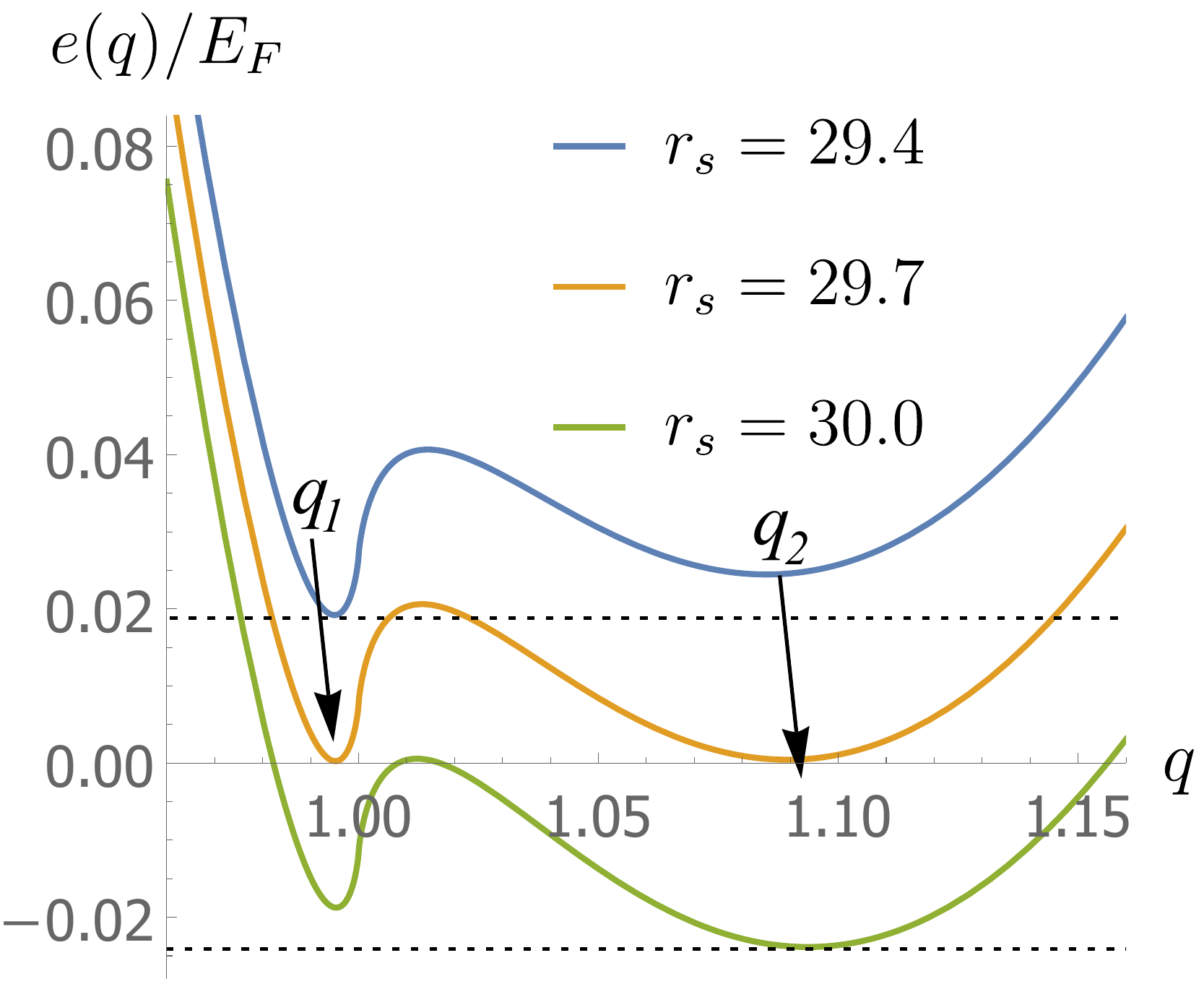}}
\subfigure[]{\label{m2400}\includegraphics[width=0.3\linewidth]{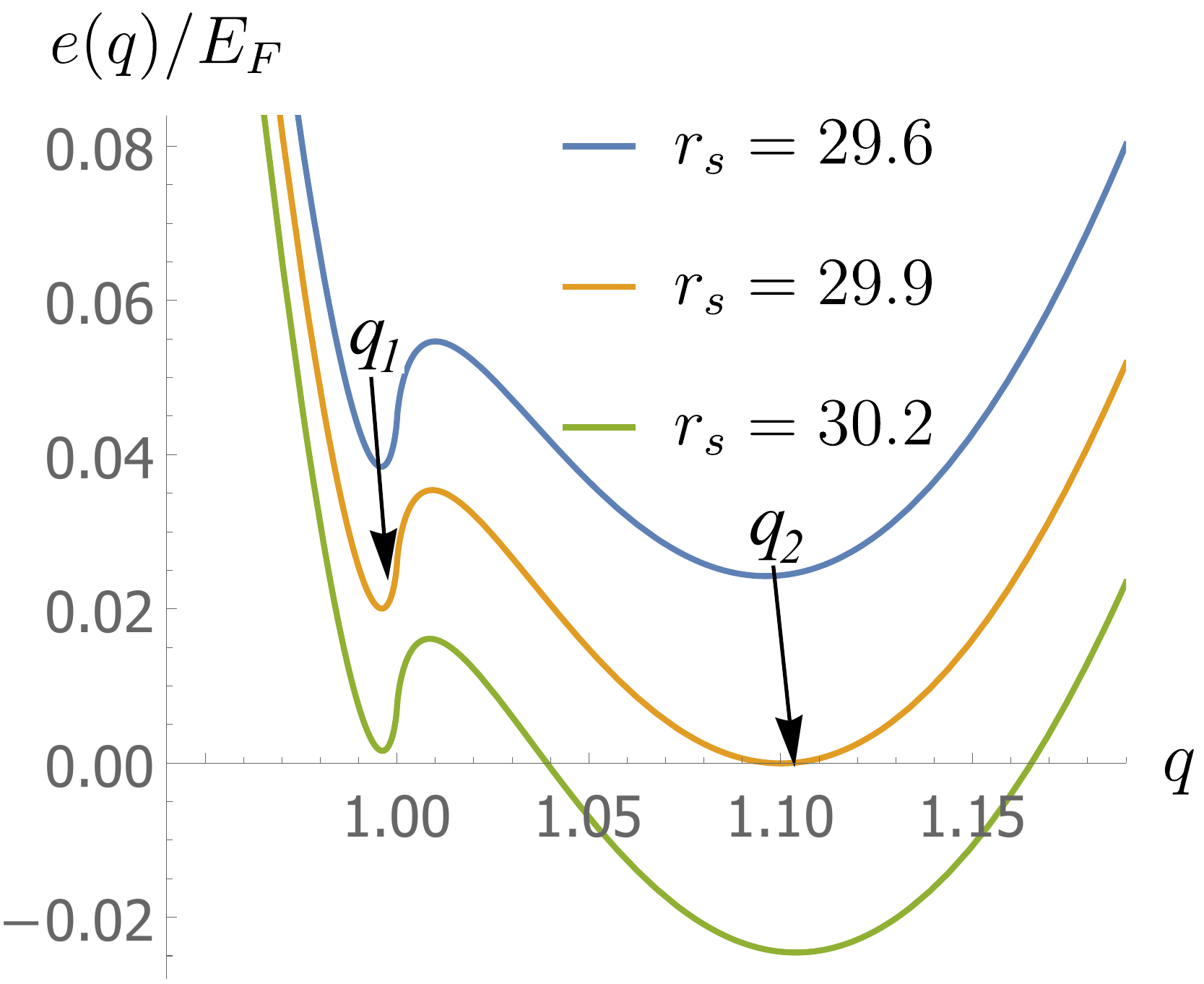}}\\
\subfigure[]{\label{m501}\includegraphics[width=0.3\linewidth]{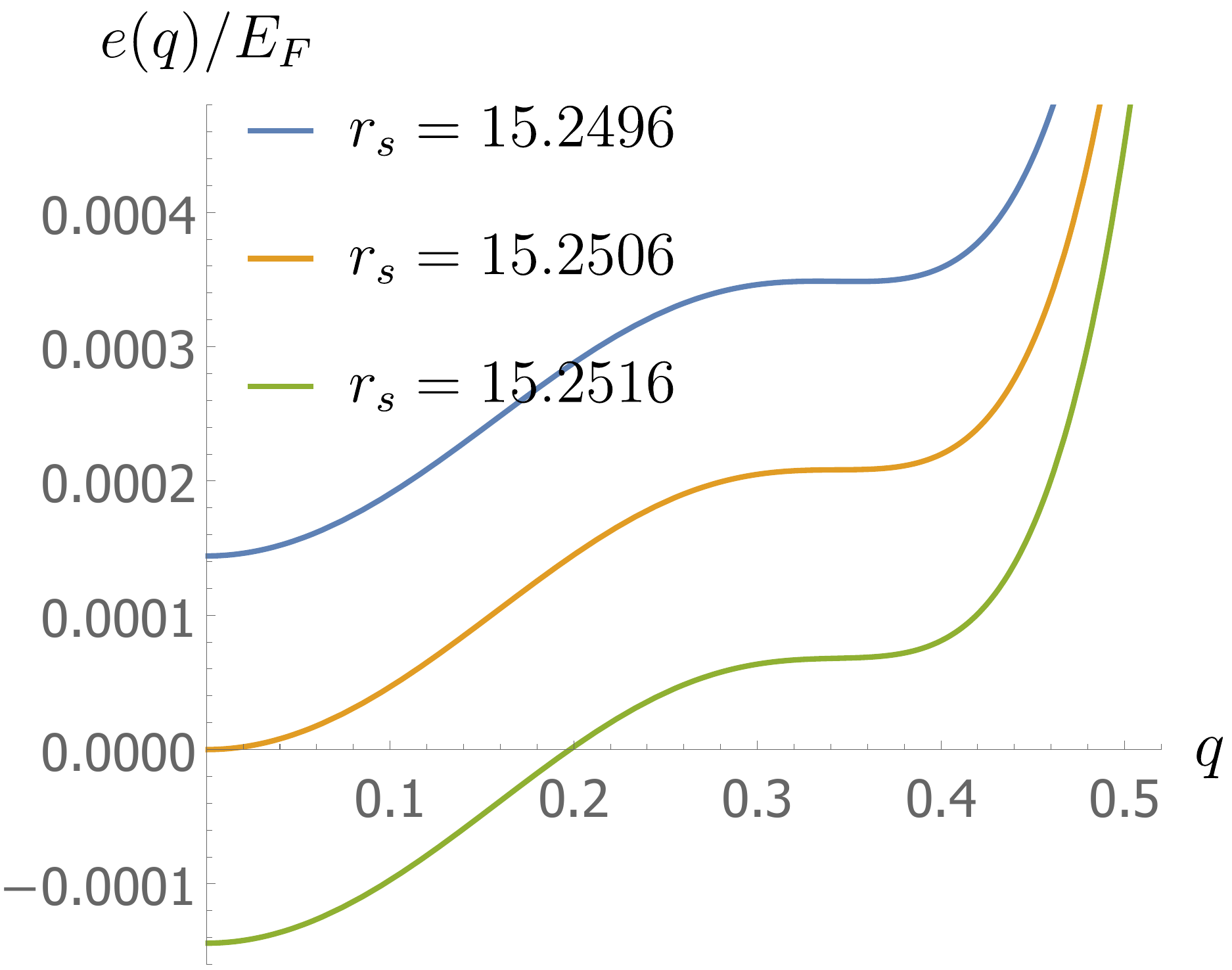}}
\subfigure[]{\label{m502}\includegraphics[width=0.3\linewidth]{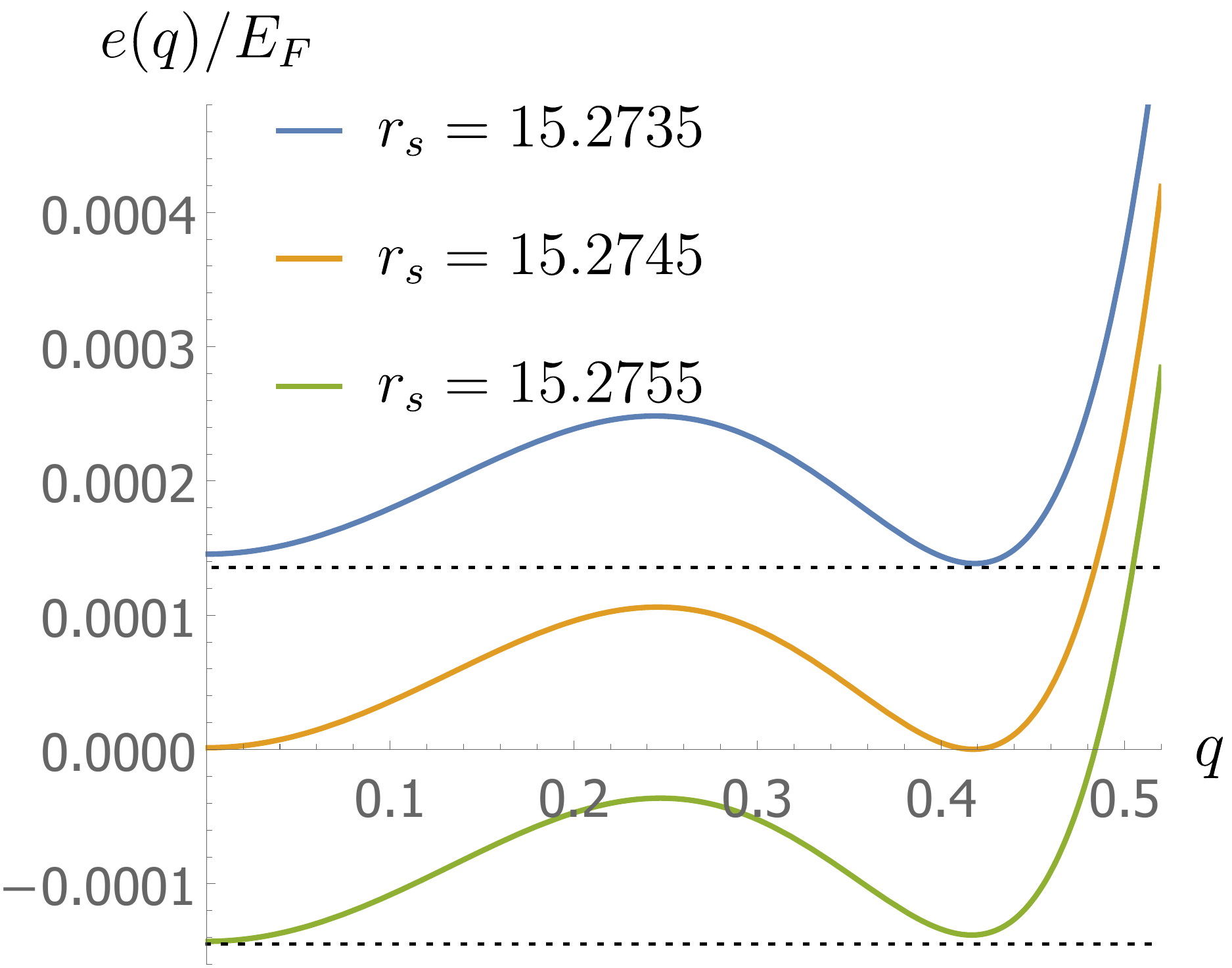}}
\subfigure[]{\label{m503}\includegraphics[width=0.3\linewidth]{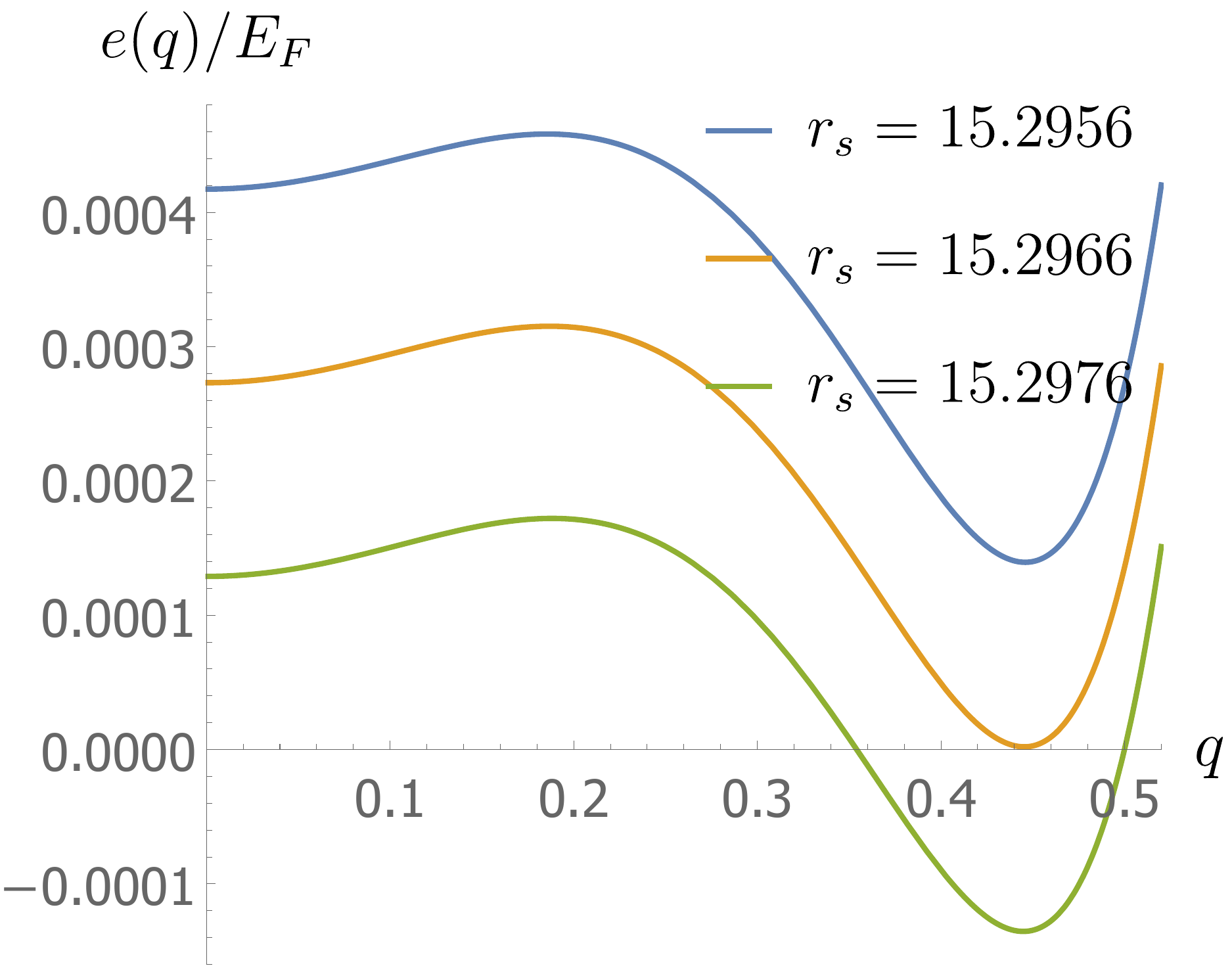}}
\caption{\label{m3} The 
energy response 
in three-dimensions
to the eigenmode with the smallest eigen-energy, 
for mass ratio $\gamma=$ (a) $24$,  (b) $240$, (c) $2400$, (d) $5.012$, (e) $5.022$ and (f) $5.032$. 
The response shown is properly normalized by the total number of particles $N$ and plotted in units of heavier particle's Fermi energy $E_F=\frac{k_F^2}{2m_*}$.
The dashed lines mark the global minima of the spectra.}
\end{figure*}
\begin{figure*}[t!] 
\subfigure[]{\label{m9}\includegraphics[width=0.3\linewidth]{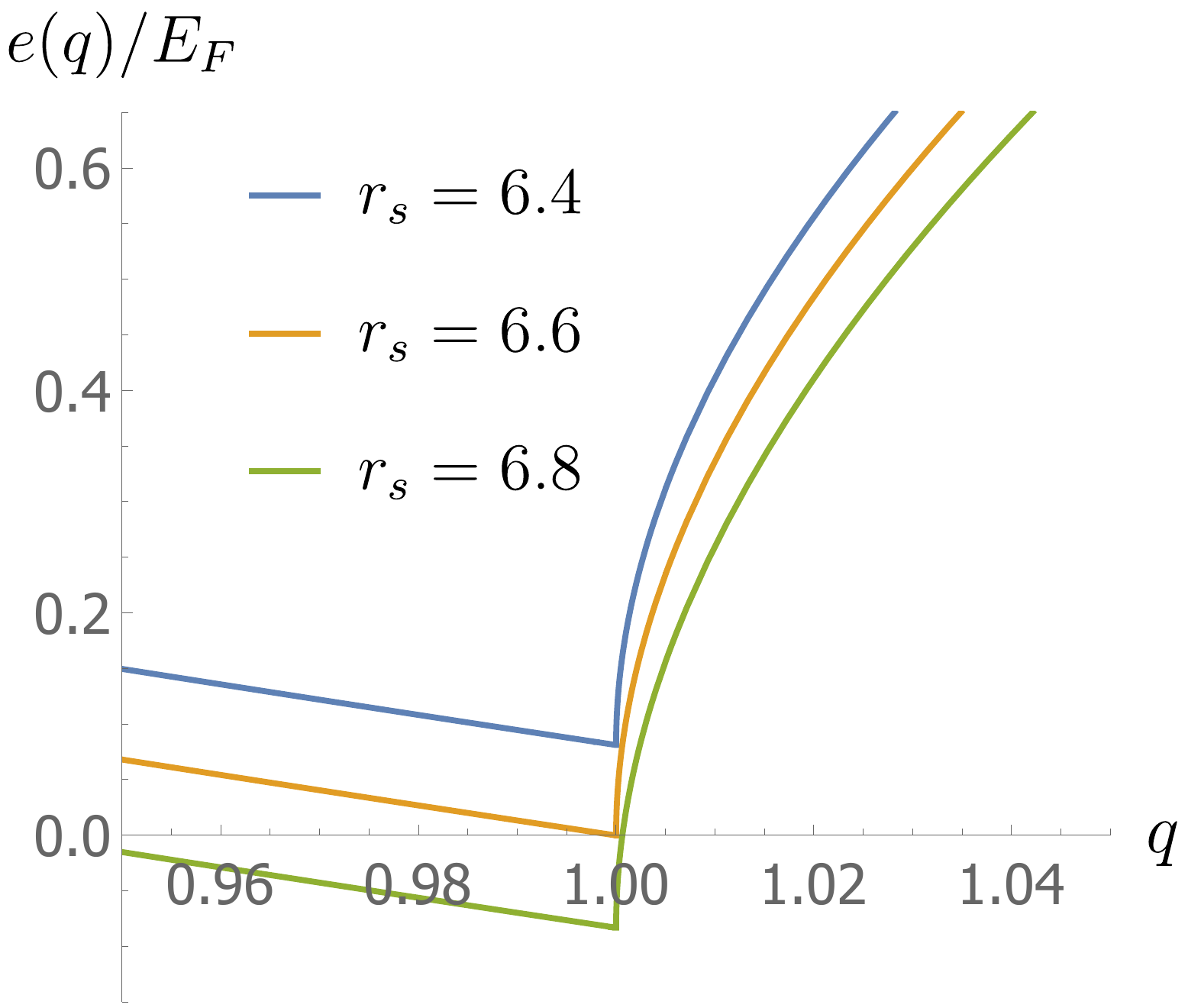}}
\subfigure[]{\label{m91}\includegraphics[width=0.3\linewidth]{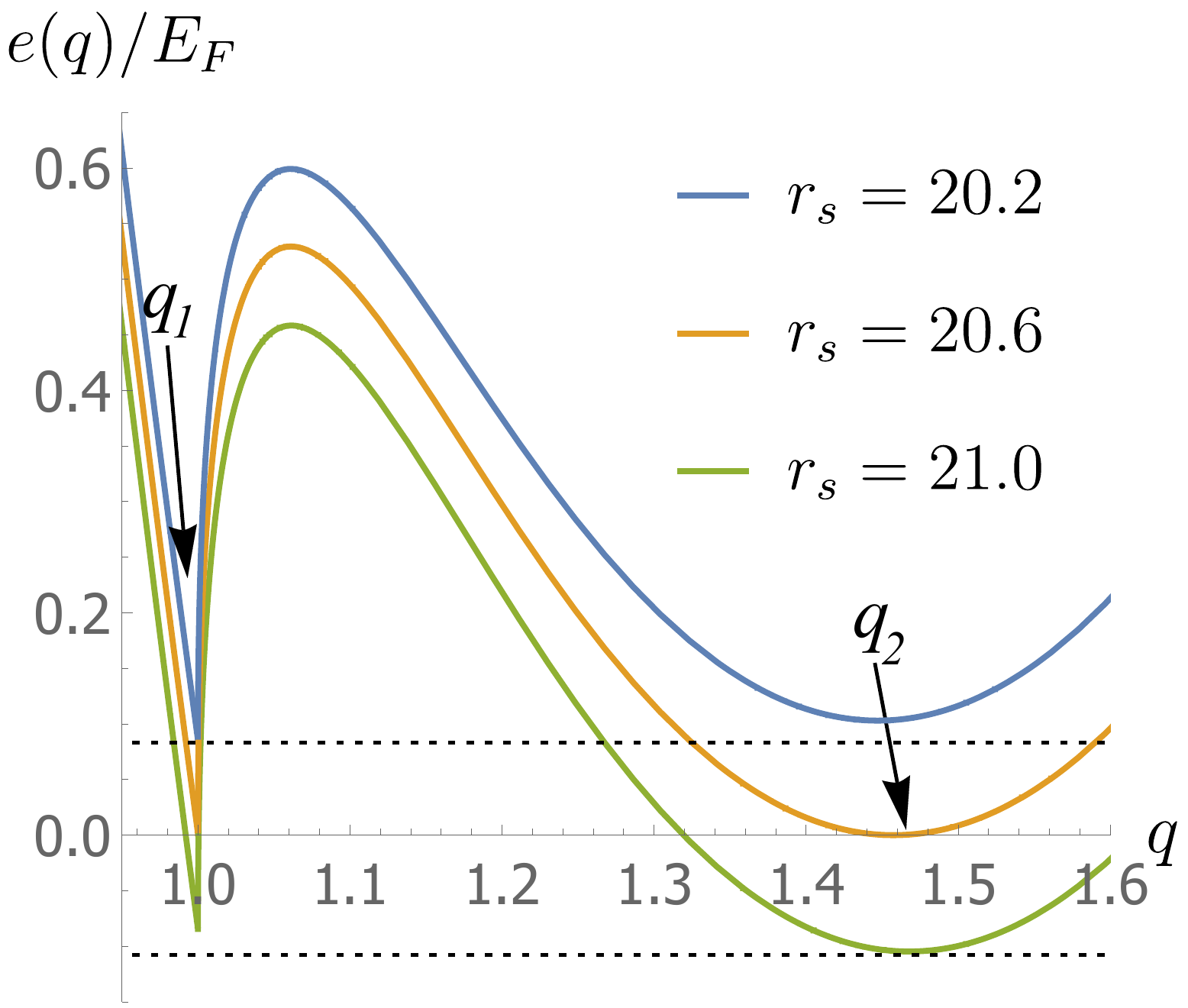}}
\subfigure[]{\label{m912}\includegraphics[width=0.3\linewidth]{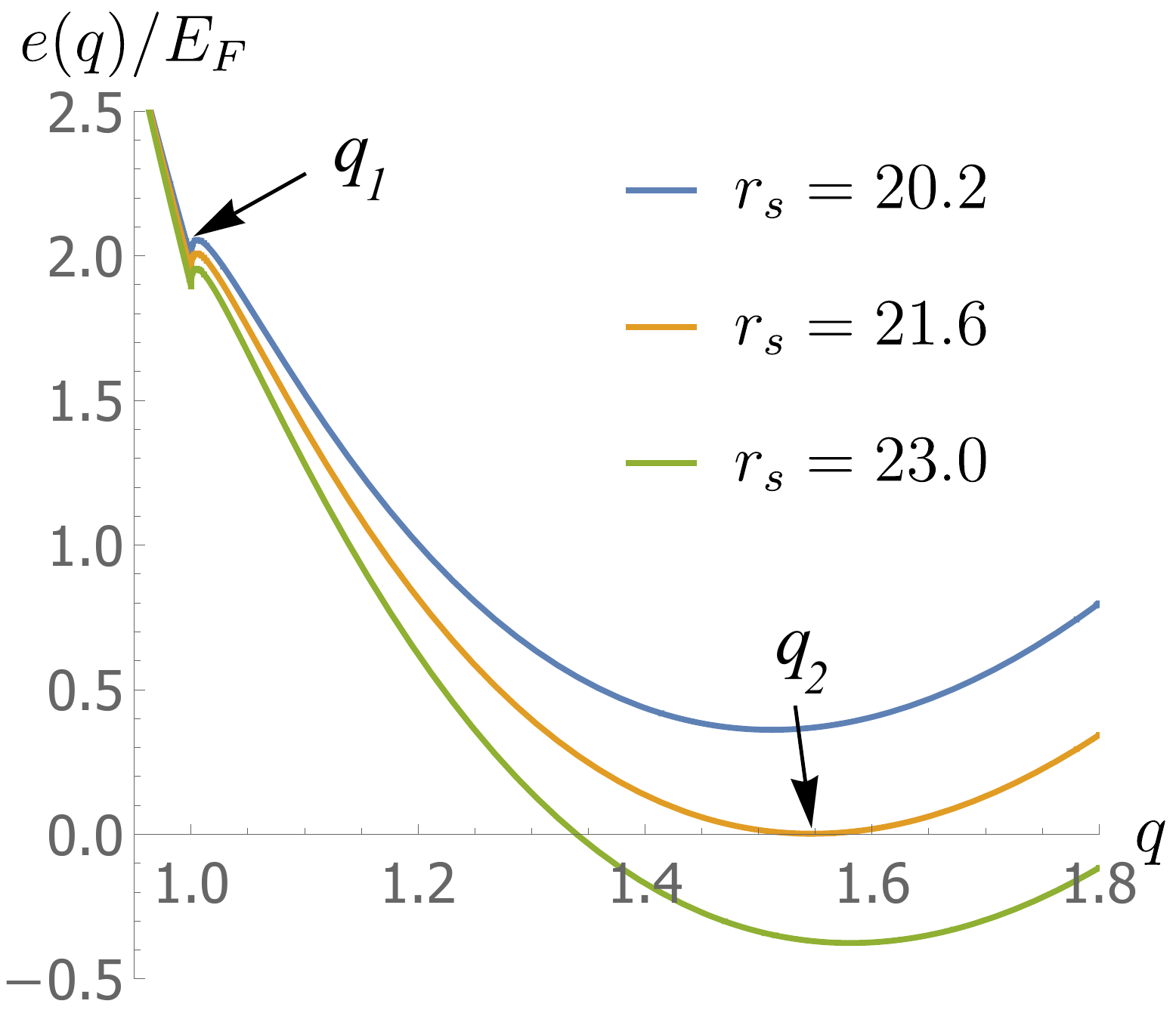}}
\caption{\label{m2} The same as in Fig.~\ref{m3}, but for the two-dimensional case, with $\gamma=$ (a) $9.12$,  (b) $91.2$ and (c) $912$.}
\end{figure*}

 In our numerical calculations of CDW spectra, 
we found no qualitative or quantitatively significant difference of the results when using different fittings of exchange-correlation energy, for example PZ81~\cite{PZ81}, PW92~\cite{PW92}, 
or SPS10~\cite{SP10}. 
We hence adopted two 
recent simple forms 
of the exchange-correlation energy in two- and three-dimensional systems~\cite{2DXC,Chachiyo}. 
In Figs.~\ref{m3} and \ref{m2}, we plot several typical energy spectra (normalized by total number of particles) of the lower-energy CDW mode, $e(q)\equiv\lambda_{\text{CDW}}^{-}(q)/N$, for different $r_s$ and $\gamma$.

We notice that each SDW mode only involves the density modulation of only one component, since there is no interaction concerning spins between the two components. In three-dimensional systems, $\lambda^{a= p, n}_{\text{SDW}}(q)$ is monotonically increasing, so the $q=0$ mode must be the first unstable SDW mode, which corresponds to a spontaneous polarization 
of the uniform $a$-gas. Moreover, if we define a polarization parameter $\eta\equiv\frac{\rho_{\uparrow}-\rho_{\downarrow}}{\rho_{\uparrow}+\rho_{\downarrow}}$, then $\lambda^{a= p, n}_{\text{SDW}}$ is proportional to the second derivative of the total energy of the uniform $a$-gas with respect to $\eta$. As shown in Fig.5 of Ref.~\cite{Zong}, past QMC results suggest that this derivative would not be negative for the three-dimensional electron gas until $r_s$ becomes larger than $\sim50$. The scaling relation in the appendix further puts this point to 
$50m_e/m_a$
for the $a$-gas. Such values of $r_s$ for polarization are much larger than the critical $r_s$ of the CDW mode analyzed below, so the unpolarized uniform plasma would first be unstable against a CDW mode and transit to a crystalline phase in our calculations. Similar argument applies in the two-dimensional case, where we again find that the CDW instability occurs earlier than the polarization point predicted by all recent studies~\cite{uni,2DXC,att02,att03}.

\begin{figure}[t!] 
\subfigure[]{\label{q0}\includegraphics[width=0.39\linewidth]{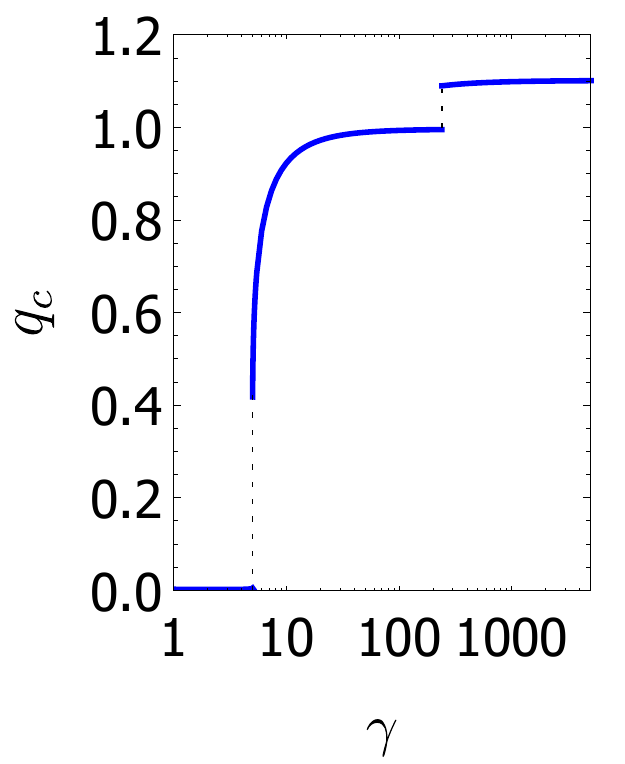}}
\subfigure[]{\label{phasediag1}\includegraphics[width=0.595\linewidth]{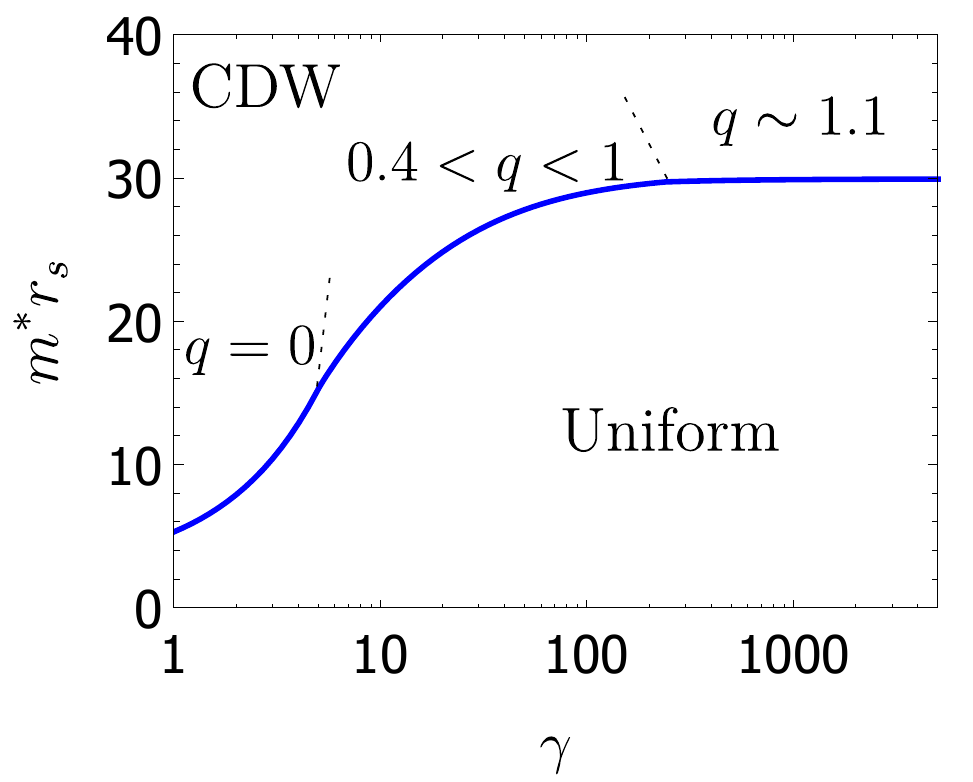}}

\caption{\label{rs3} 
(a) The first unstable CDW wave vector $q_c$ when decreasing density for different mass ratios and 
(b) the phase diagram of the plasma, 
in three dimensions.
The solid line in (b) 
is the exact critical $r_s$ - $\gamma$ 
relation, normalized by the mass scale $m^*$. The dashed lines are conjectured from the $r_s$-dependence of the energy spectra.}

\end{figure}

\begin{figure}[t!] 
\subfigure[]{\label{q02}\includegraphics[width=0.39\linewidth]{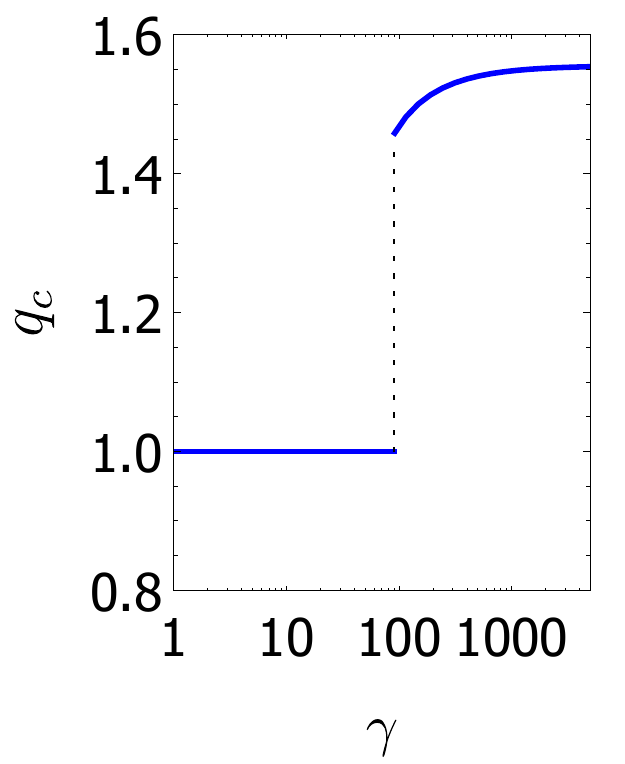}}
\subfigure[]{\label{phasediag2}\includegraphics[width=0.595\linewidth]{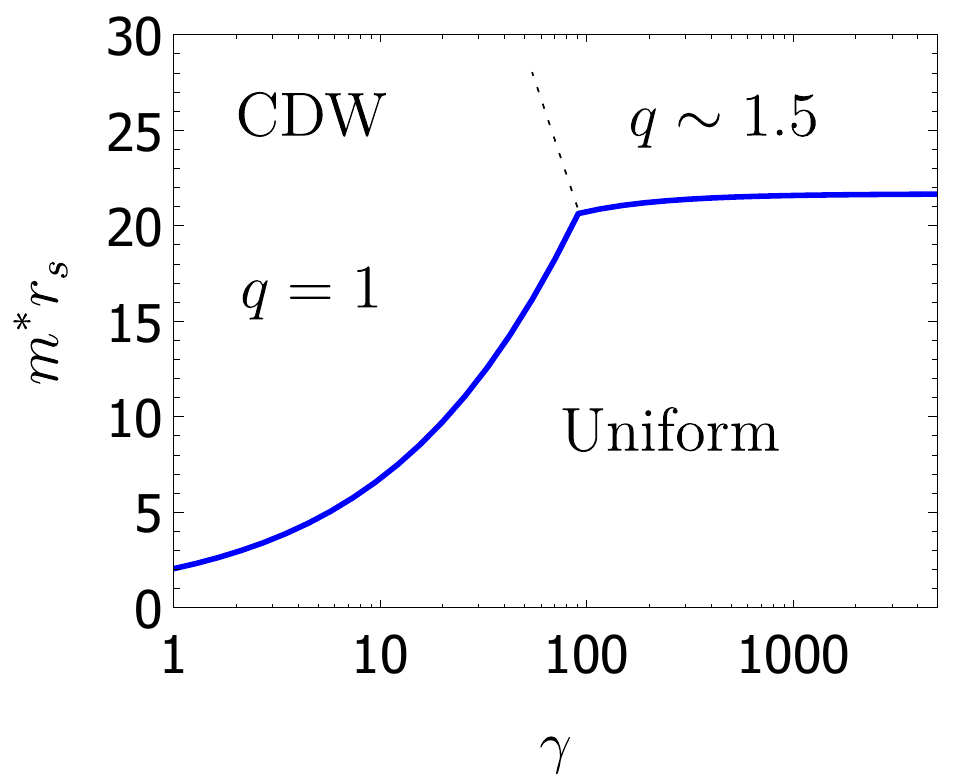}}
\caption{\label{rs2} The same as in Fig.~\ref{rs3}, but for 
2D. 
}

\end{figure}

Now we turn to the analysis on the CDW mode whose eigen-energy first approaches zero as increasing $r_s$ and its corresponding wavevector $q_c$. It is worth noting that, due to the Kohn anomaly of the Lindhard function, the CDW spectra share a positively divergent gradient at $q=1$. Thus for certain $r_s$ and $\gamma$ 
a downtrend can be introduced near that point, and the spectra would show 
a double well structure,
as shown in Figs.~\ref{m240} and \ref{m2400} for 3D, and Figs.~\ref{m91} and \ref{m912} for 2D, where we have labeled the 
two local minima on both sides of $q=1$ by $q_{1}$ and $q_{2}$.
This fine structure introduces discontinuities of the leading symmetry breaking wave-vector as 
$\gamma$ is varied. We note that this is a 
pure quantum effect which is related to the Fermi surface and thus the Pauli exclusion principle. Similar but finer structures in the spectra also 
occur when $m_p$ is close to $m_n$ for 
3D systems, as shown in Figs.~\ref{m501}-\ref{m503}. These sudden changes of the wavelength of the leading unstable CDW eigenmode  
is seen more clearly in Fig.~\ref{q0},  
where we plot the wave vector length $q_c$ of the first unstable CDW wave 
against the mass ratio $\gamma$.
A similar plot is shown for 2D in Fig.~\ref{q02}.

To conclude these results, we plot the critical value $r_s^c$, which is normalized by the mass scale $m^*$, versus the mass ratio $\gamma$ 
in Figs.~\ref{phasediag1} and \ref{phasediag2} for 3D and 2D respectively. These can be viewed as phase diagrams indicating the transition line between different crystalline phases and a uniform liquid phase. 
The line can
be divided into parts and corresponds to different intervals of leading symmetry breaking wave-vector, which may indicate exotic quantum structural phase transitions between different crystalline phases with discontinuous lattice constant. 
Furthermore, 
from Figs.~\ref{m240} and \ref{m502} in 3D and Fig.~\ref{m91} in 2D,
we can see that changing $r_s$ while fixing $\gamma$ can also alter the choice of global minimum between the two local minima in the spectrum.
Based on this information we can infer possible phases in the vicinity of 
the phase transition line, as we have indicated with the dashed lines
in Figs.~\ref{phasediag1} and \ref{phasediag2}.

\section{Discussions}\label{dis}

We first 
discuss two interesting limiting cases. 
The first is the jellium limit as $\gamma\rightarrow \infty$ and $m^* = 1$. In this case, the heavier component becomes the electrons and the lighter one is so free that the only role it could play is a uniform background.
In three dimensions, we obtain a critical Wigner radius $r_s\approx29.9$ and a leading unstable CDW mode of wave-vector $q_c \approx 1.10  \ (2k_F)$. In the two-dimensional case, 
the critical Wigner radius $r_s\approx21.7$ and the corresponding wave-vector $q_c \approx 1.56\ (2k_F)$. These points are close to the earlier results acquired by similar methods~\cite{Perdew,Sander}. These symmetry breaking points occur earlier than predicted by QMC.
Thus our result could be a hint for the existence of new intermediate phases for the ground state of uniform electron gas, with the
discrepancy 
arising from the possibility that the candidate structures
searched in QMC calculations so far are not yet optimal.
However, it is also likely the result of the approximate treatment
of the original many-body Hamiltonian
by a density-functional under LSDA, especially since this is in the regime of large 
$r_s$ with strong correlation effects.
Our predicted symmetry-breaking points are later than those from HF.
This is also reasonable 
since HF 
will consistently overestimate the trend of (especially magnetic) inhomogeneity. 

A closely related case is hydrogen, with $\gamma=m^\star \approx 1837$.
Our results suggest the onset of CDW at a tiny $r_s\sim 0.016$ (a.u.). 
(We should note that at such high densities, relativistic effects are important, which are not accounted for in our theory.)
This indeed corresponds to 
a much higher density than the regime where 
previous more detailed calculations ~\cite{chen2013quantum,RevModPhys.84.1607,1989ASIB..186..477C}
have identified atomic orders. Our result may shed light on the possibility of quantum solid phase in hydrogen at an ultra high density which is far beyond reach of today's experiment. 

The second case is the $1\leq\gamma\leq5.022$ region 
in 3D, where the leading unstable wave-vector is $0$. Near the critical point, the $q\rightarrow0$ modes of negative energies have $\bm{\rho}^{-}_{\text{CDW}} \approx \left(1,1,1,1\right)/2$, which indicates that the system favors to 
bodily move and the macro fluctuations induced by the long-wavelength density waves would force the system to be no longer confined by the volume $V$ but self-confined to a denser state by cohesive force. 
In other words, phase separation occurs in this regime, which 
also implies the nonexistence of quantum crystal when the masses of the two components are close. Similar results have also been reported at finite temperatures ~\cite{DPIMC2003A, DPIMC2003B, CrytwoCom}.

As mentioned, it is a major approximation to treat the $p$-$n$ interaction 
only by the Hartree approximation. 
It is reasonable to consider whether adding back the $p-n$ correlation effect would change our results. This of course can not be definitely answered 
without a better treatment. 
However a few hints are available from formal considerations. 
For example we could consider 
adding back
a $p$-$n$ correlation functional $E^{p{\rm -}n}_{\text{c}}$ in Eq.~\ref{functional} like in a previous DFT study on the two component system~\cite{Nieminen1986}. This term would modify the behavior of the mediating electronic force at second order of expansion, but would not 
eliminate the Kohn anomalies in the spectra,  thus not the double well structures near $q=1$ for intermediate mass ratios. Moreover, it can be easily checked that the critical $q_c$ remains unchanged at the two limiting cases, 
$\gamma\rightarrow1$ and $\infty$. 
This would imply that,
as $\gamma$
varies from $1$ to $\infty$, $q_c$ still must go through a discontinuity around $1$. Thus we 
conclude that at least the correlation effect would not qualitatively affect the existence of the discontinuity near $2k_F$ in the relation between critical wave-vector and mass ratio. We remark that, since the anomalous discontinuity is rooted from the nature of fermionic response functions and varying mass ratio is equivalently tuning the strength of Coulomb screening, then adjusting other parameters that plays the same role might also introduce similar phenomena in different system settings.

There are indeed regimes where this framework breaks down. For example when 
$\gamma=1$, QMC calculations indicate that Bose condensation of exitonic molecules occurs at rather small $r_s$~\cite{shumway2000path,QMCeh}. In these situations, we believe the following generalization of our approach would lead to significant improvements while adding little additional complexity. 
We could consider a Kohn-Sham variational wave function in the form of 
a product of projected BCS wave functions (Antisymmetrized Germinal Powers, AGPs), each of which describes a pairing state between the two species 
(for example, one for pairing between $p_\uparrow$ and $n_\downarrow$, while the other for  $p_\downarrow$ and $n_\uparrow$). 
The computational manipulations necessary for using such a wave function with the Kohn-Sham plus p-n Hartree Hamiltonian 
are readily available (see e.g., Ref.~\cite{HaoShiHFB}). 

It is worth noting that, if the energy functional remains 
valid at small density variation, 
the exotic phase transition around $\sim 2k_F$ could also be identified by probing different energy dispersion relations of the phonon-like Goldstone mode. Expanding a spatially slowly 
varying phase $u(\bm{x})$ of the condensed amplitude $\bm{\rho}_{\bm{q}_c}=|\bm{\rho}_{\bm{q}_c}|e^{-\mathrm{i}u(\bm{x})}$, we find that the energy dispersion of the $u_{\bm{p}}$ mode is proportional to that of the $\bm{\rho}_{\bm{q}_c+\bm{p}}$ mode. Thus the quantitative (for 3D) or qualitative (for 2D) difference in the appearance of the energy dispersions around two local minima, shown in Figs.~\ref{m240} for 3D and \ref{m91} for 2D, may possibly be observed by spectroscopic experiments. This is especially interesting for the case of the minimum lying exactly at $2k_F$ in the 2D system. The sharp turning of the CDW mode dispersion indicates a linear (quadratic) dispersion of the Goldstone mode along (perpendicular to) the symmetry breaking direction, which is different 
from traditional theory of the elastic behavior for short-range correlated smectic liquid crystals ~\cite{chaikin}.

Lastly, we remark that our results suggest the possible existence of ``quantum crystals.''  
Since tuning parameters such as 
mass ratio can change the characteristic length scale in the system, lattices at intermediate density can possess non-integer numbers of particles per unit cell, which is never the case in classical crystals. This is an interesting direction for further investigations, for example with more explicit calculations.


\section{Conclusion}
In summary,
we have proposed a quantum model for a two-component fermionic plasma and a theoretical approach for treating it. We formulate an approximate numerical solution based on the theory of DFT using LSDA, and obtain the critical values of density and wave-vector where an instability of the uniform state against a CDW occurs. When the mass ratio is varied from $1$ to $+\infty$, we identify several distinct ranges of critical CDW wave-vector lengths in both the two- and three-dimensional cases, which may indicate different structures of quantum crystalline phases. Zero-temperature phase diagrams are provided. A simple scaling relation is given which 
allows the results to be generalized to any mass scale. With the framework presented in this work, one can expect that higher order perturbative expansions of the functional would support the analysis on possible instabilities towards more exotic density ordering phases (e.g. non-collinear magnetism).

\section*{Acknowledgement}

We thank Steven A.~Kivelson, David M.~Ceperley, Markus Holzmann, and Xin-Zheng Li for helpful discussions.
S.Z.~acknowledges support from NSF DMR-1409510.
The Flatiron Institute is a division of the Simons Foundation.

\bibliographystyle{apsrev4-1} 
\bibliography{plasma}

\appendix

\section{A Scaling Relation for Different Mass Scales}\label{a}
We have the Hamiltonian
\begin{align}
\hat{H}=-\sum_i\frac{1}{2m_{i}}\nabla_{i}^2+\sum_{i<j} \frac{q_iq_j}{|\bm{r_i}-\bm{r_j}|}
\end{align}
for a system confined in a given $D$-dimensional volume $V$ consisting of several components of $N$ charged fermions. 
A Wigner radius can still be defined, as in the main text, to parametrize the number density $\rho_0=N/V$. Let $\bm{\xi} =(\bm{r}_1,\bm{r}_2,...\bm{r}_N)$ and suppose that $\psi(\bm{\xi})$ is an eigenstate of such a system, which satisfies:
\begin{align}
\hat{H}\psi(\bm{\xi})=E\psi(\bm{\xi})
\end{align}

Now we perform a coordinate transformation, by replacing all $\bm{r}_i$ by $k\bar{\bm{r}}_i$ in the equation above:
\begin{align}
\left(-\sum_i\frac{1}{2m_{i}k^2}\bar{\nabla}_{i}^2+\sum_{i<j} \frac{q_iq_j}{k|\bar{\bm{r}}_i-\bar{\bm{r}}_j|}\right)\psi(k\bar{\bm{\xi}})=E\psi(k\bar{\bm{\xi}})\,.
\end{align} 
Defining a compressed wave-function $\bar{\psi}(\bar{\bm{\xi}}) \equiv \psi (k\bm{\xi})$ in the space $\bar{V} = V / k^D$, and rearranging the equation into the form:
\begin{align}
\left(-\sum_i\frac{1}{2km_{i}}\bar{\nabla}_{i}^2+\sum_{i<j} \frac{q_iq_j}{|\bar{\bm{r}}_i-\bar{\bm{r}}_j|}\right)\bar{\psi}(\bar{\bm{\xi}})=kE\cdot\bar{\psi}(\bar{\bm{\xi}})\,,
\end{align}
we can immediately see that the operator on the left-hand side is the Hamiltonian  of another system with each component having $k$ times larger mass, i.e. $\bar{m}_{i} = km_i$. Hence we know that $\bar{E}\equiv kE$ is an eigen-energy and $\bar{\psi}(\bar{\bm{\xi}})$ is the corresponding eigenstate of the new system confined in the volume $\bar{V} = V / k^D$ with $\bar{r}_s=r_s/k$. We mention that a similar argument has been proposed in ~\cite{mass} and that these relations 
can be thought of as applications of more general scaling theory in the renormalization group~\cite{Altland}.

In particular, this system can be a two-component plasma and the eigenstate can be the ground state. Thus the ground-state energy and density distributions we have obtained in this paper can be easily generalize to all combinations of the masses, by scaling the whole system.

For the case of one-component electron system, the scaling relations of the interacting and non-interacting uniform ground-states read:
\begin{align}
E_0(r_s) = k \bar{E}_0(kr_s) \nonumber\\
T_0(r_s) = k \bar{T}_0(kr_s) \nonumber
\end{align}
where $T_0$ and $\bar{T}_0$ are the kinetic energies within the Fermi spheres and $E_0$ and $\bar{E}_0$ are the uniform ground-state energies of two jellium systems with mass $m_e$ and $km_e$. 
Recalling the definition of exchange-correlation energy of uniform electron gas, $\epsilon_{\text{xc}}=(E_0-T_0)/N$, we acquire the exchange-correlation energy $\bar{\epsilon}_{\text{xc}}(\bar{r}_s)=k \epsilon_{\text{xc}}(k \bar{r}_s)$ for particles with mass $km_e$ and equal charge. We note that this scaling relation holds for any polarization since the scaling operation does not change the ratio between up- and down-spin particles.

\end{document}